\documentclass[twocolumn,trackchanges]{aastex701}
\usepackage{enumitem}
\usepackage{hyperref}
\usepackage[version=4]{mhchem}
\usepackage{amsmath}
\usepackage{natbib}
\usepackage{subcaption}
\usepackage{graphicx}
\usepackage{caption}
\usepackage{xcolor}

\begin{document}

\title{WFST Supernovae in the First Year: III. Systematical Study of the Photometric Behavior of Early-phase Core-collapse Supernovae}

\author{Junhan Zhao}
\affiliation{Department of Astronomy, University of Science and Technology of China, Hefei 230026, China}
\affiliation{School of Astronomy and Space Sciences, University of Science and Technology of China, Hefei, 230026, China}
\email{zjunhan@mail.ustc.edu.cn}

\author[0000-0002-9092-0593]{Ji-an Jiang}
\affiliation{Department of Astronomy, University of Science and Technology of China, Hefei 230026, China}
\affiliation{National Astronomical Observatory of Japan, 2-21-1 Osawa, Mitaka, Tokyo 181-8588, Japan}
\textcolor{blue}{\email{jian.jiang@ustc.edu.cn}}

\author{Zelin Xu}
\affiliation{Department of Astronomy, University of Science and Technology of China, Hefei 230026, China}
\affiliation{School of Astronomy and Space Sciences, University of Science and Technology of China, Hefei, 230026, China}
\email{sa21022025@mail.ustc.edu.cn}

\author[0000-0002-1067-1911]{Yu-Hao Zhang}
\affiliation{Institute of Astrophysics, Central China Normal University, Wuhan 430079, China; }
\affiliation{Laboratory for Compact Object Astrophysics and Astronomical Technology, Central China Normal University, Wuhan 430079, China}
\affiliation{Education Research and Application Center, National Astronomical Data Center, Wuhan 430079, China}
\email{zhang-yh@mails.ccnu.edu.cn}

\author{Qiliang Fang}
\affiliation{National Astronomical Observatory of Japan, 2-21-1 Osawa, Mitaka, Tokyo 181-8588, Japan}
\email{qiliang.fang@nao.ac.jp}

\author[0000-0002-8708-0597]{Liang-Duan Liu}
\affiliation{Institute of Astrophysics, Central China Normal University, Wuhan 430079, China; }
\affiliation{Laboratory for Compact Object Astrophysics and Astronomical Technology, Central China Normal University, Wuhan 430079, China}
\affiliation{Education Research and Application Center, National Astronomical Data Center, Wuhan 430079, China}
\email{liuld@ccnu.edu.cn}

\author{Qingfeng Zhu}
\affiliation{Department of Astronomy, University of Science and Technology of China, Hefei 230026, China}
\affiliation{Institute of Deep Space Sciences, Deep Space Exploration Laboratory, Hefei 230026, China}
\textcolor{blue}{\email{zhuqf@ustc.edu.cn}}

\author[0000-0002-1067-1911]{Yun-Wei Yu}
\affiliation{Institute of Astrophysics, Central China Normal University, Wuhan 430079, China; }
\affiliation{Laboratory for Compact Object Astrophysics and Astronomical Technology, Central China Normal University, Wuhan 430079, China}
\affiliation{Education Research and Application Center, National Astronomical Data Center, Wuhan 430079, China}
\email{yuyw@ccnu.edu.cn}

\author[0000-0003-2611-7269]{Keiichi Maeda}
\affiliation{Department of Astronomy, Kyoto University, Kitashirakawa-Oiwake-cho, Sakyo-ku, Kyoto 606-8502, Japan}
\email{keiichi.maeda@kusastro.kyoto-u.ac.jp}

\author{Lluís Galbany}
\affiliation{Institute of Space Sciences (ICE-CSIC), Campus UAB, Carrer de Can Magrans, s/n, E-08193 Barcelona, Spain}
\affiliation{Institut d'Estudis Espacials de Catalunya (IEEC), 08860 Castelldefels (Barcelona), Spain}
\email{lluisgalbany@gmail.com}

\author{Hanindyo Kuncarayakti}
\affiliation{Tuorla Observatory, Department of Physics and Astronomy, FI-20014 University of Turku, Finland}
\affiliation{Finnish Centre for Astronomy with ESO (FINCA), FI-20014 University of Turku, Finland}
\email{kuncarayakti@gmail.com}

\author{\v{Z}eljko Ivezi\'{c}}
\affiliation{Department of Astronomy, University of Washington, Box 351580, Seattle, Washington 98195-1580, USA}
\email{ivezic@uw.edu}

\author{Saurabh W. Jha}
\affiliation{Department of Physics and Astronomy, Rutgers, The State University of New Jersey, 136 Frelinghuysen Road, Piscataway, New Jersey 08854, USA}
\email{saurabh@physics.rutgers.edu}

\author{Peter Yoachim}
\affiliation{Department of Astronomy, University of Washington, Box 351580, Seattle, Washington 98195-1580, USA}
\email{yoachim@uw.edu}

\author{Dezheng Meng}
\affiliation{Department of Astronomy, University of Science and Technology of China, Hefei 230026, China}
\affiliation{School of Astronomy and Space Sciences, University of Science and Technology of China, Hefei, 230026, China}
\email{dezhengmeng@mail.ustc.edu.cn}

\author{Weiyu Wu}
\affiliation{Department of Astronomy, University of Science and Technology of China, Hefei 230026, China}
\affiliation{School of Astronomy and Space Sciences, University of Science and Technology of China, Hefei, 230026, China}
\email{wwyyy@mail.ustc.edu.cn} 

\author{Zhengyan Liu}
\affiliation{Department of Astronomy, University of Science and Technology of China, Hefei 230026, China}
\email{ustclzy@mail.ustc.edu.cn}

\author{Andrew J. Connolly}
\affiliation{Department of Astronomy, University of Washington, Box 351580, Seattle, Washington 98195-1580, USA}
\email{ajc@astro.washington.edu}

\author{Ziqing Jia}
\affiliation{Department of Astronomy, University of Science and Technology of China, Hefei 230026, China}
\email{zqjia@mail.ustc.edu.cn}

\author{Wen Zhao}
\affiliation{Department of Astronomy, University of Science and Technology of China, Hefei 230026, China}
\email{wzhao7@ustc.edu.cn}

\author{Lulu Fan}
\affiliation{Department of Astronomy, University of Science and Technology of China, Hefei 230026, China}
\affiliation{Institute of Deep Space Sciences, Deep Space Exploration Laboratory, Hefei 230026, China}
\email{llfan@ustc.edu.cn}

\author{Ming Liang}
\affiliation{National Optical Astronomy Observatory (NSF’s National Optical-Infrared Astronomy Research Laboratory) 950 N Cherry Ave. Tucson Arizona 85726, USA}
\email{liangming@gmail.com}

\author{Hairen Wang}
\affiliation{Purple Mountain Observatory, Chinese Academy of Sciences, Nanjing 210023, China}
\email{hairenwang@pmo.ac.cn}

\author{Jian Wang}
\affiliation{State Key Laboratory of Particle Detection and Electronics, University of Science and Technology of China, Hefei 230026, China}
\affiliation{Institute of Deep Space Sciences, Deep Space Exploration Laboratory, Hefei 230026, China}
\email{wangjian@ustc.edu.cn}

\author{Hongfei Zhang}
\affiliation{State Key Laboratory of Particle Detection and Electronics, University of Science and Technology of China, Hefei 230026, China}
\email{nghong@ustc.edu.cn}

\correspondingauthor{Ji-an Jiang, Qingfeng Zhu}
\email{jian.jiang@ustc.edu.cn, zhuqf@ustc.edu.cn}

\begin{abstract}
We investigate the multiband photometric properties of seven supernovae (SNe) showing double-peaked light-curve evolution and prominent shock-cooling emission, observed by the Wide Field Survey Telescope (WFST) during its first year of operation. By jointly employing an analytic early shock-cooling model and the Arnett radioactive-diffusion model, we fit the bolometric light curves and infer ejecta masses in the range $1.1$–$2.6 M_\odot$, consistent with a transitional population between ultra-stripped supernovae (USSNe) and normal stripped-envelope supernovae (SESNe). The envelope masses are estimated to be $M_{\rm env}=0.1$–$0.4 M_\odot$, while the progenitors are constrained to be yellow or blue supergiants (YSGs/BSGs) with radii of $R=120$–$300 R_\odot$. Using empirical relations, we estimate progenitor luminosities of $L=10^{4.6}$–$10^{4.9} L_\odot$, corresponding to zero-age main-sequence (ZAMS) masses of $8$–$20 M_\odot$. Theoretical models suggest that such progenitors are more naturally produced through binary evolution channels, as single-star evolutionary pathways are unable to yield ejecta masses this low.

\end{abstract}


\section{Introduction}

Stellar evolution theory predicts that massive stars ($\gtrsim 8\,M_\odot$) experience core collapse at the end of their lives, exploding as core-collapse supernovae (CCSNe) and resulting in the formation of neutron stars or black holes. \citep{gal-yam_observational_2017, janka_explosion_2012}. Stripped-envelope supernovae (SESNe) constitute a major subset of CCSNe, representing the explosions of massive stars that have lost most or all of their H/He envelopes by the end of their lives due to binary interactions, stellar winds, or other physical mechanisms \citep{yoon_type_2017, fang_hybrid_2019, sun_uv_2023}. Depending on the extent of envelope stripping, SESNe are spectroscopically classified into IIb $\xrightarrow{}$ Ib $\xrightarrow{}$ Ic with increasing degrees of stripping \citep{gal-yam_observational_2017, karamehmetoglu_ogle-2014-sn-131_2017, prentice_physically_2017}.

A typical characteristic of many type IIb SNe is their double-peaked light curves (e.g., SN~1993J \citealt{woosley_sn_1994}; SN 2016gkg \citealt{bersten_surge_2018, arcavi_constraints_2017}; SN 2017jgh \citealt{armstrong_sn2017jgh_2021}; SN 2020bio \citealt{pellegrino_sn_2023}; SN 2021zby \citealt{wang_revealing_2023}; SN 2024abfo \citealt{reguitti_sn_2025}), among these, SN 2017jgh and SN 2021zby show a complete and prominent cooling peak due to the high cadence and long term monitoring of Kepler telescope and TESS, but there was no sign of shock breakout rise before shock-cooling rise probably due to the low brightness. Moreover, with the advent of wide-field surveys such as ATLAS and ZTF, an increasing number of type Ib/c SNe have been found to show double-peaked light curves. These features are interpreted as signatures of the shock traversing circumstellar material (CSM) or envelope formed during binary interactions,  wind, wave heating process or pulsational pair-instability mechanisms, followed by subsequent cooling of the ejected material (\citealt{das_probing_2023, das_probing_2024}, hereafter D23 and D24, respectively). Theoretically, the initial peak of the light curve is commonly attributed to the shock breakout (SBO), which marks the moment when the supernova (SN) shock wave -- generated by the core bounce during collapse -- breaks out from the surface of the progenitor star. This phenomenon produces a brief but intense flash of radiation, often considered as the first observable signal of a CCSN. The duration of the SBO is extremely short, typically lasting from seconds to hours \citep{waxman_shock_2017}. As a result, detecting this phase poses significant observational challenges.

Following the SBO, the early-time light curve evolution is dominated by the shock-cooling emission, which arises as the outer layers of the ejecta heated by the shock and cool down adiabatically. This phase typically manifests as a relatively rapid decline in brightness over the first few days after explosion. The detailed morphology of this early peak—including its timing, duration, and luminosity—carries valuable information about the structure and extent of the progenitor’s outer envelope \citep{piro_shock_2021}. Therefore, efforts to detect and characterize the early rise and decline of the light curve are crucial for constraining the final evolutionary stages of massive stars and improving our understanding of the pre-SN mass-loss processes. In this paper, we investigate SNe with early-excess emission discovered in the first year of the Wide Field Survey Telescope (WFST) operation, fitting their light curves to constrain progenitor properties and possible evolutionary origins.

This paper is organized as follows. The observation and data reduction are presented in Section \ref{sec:obs}, followed by the analysis in Section \ref{sec:analysis}, where photometric properties are analyzed. The analytical modeling of the early phase light curves and the explosion parameters are presented in Section \ref{sec:discussion}, and finally, the results are discussed and summarized in Section \ref{sec:conclusion}.

\begin{table*}
\caption{Properties of WFST SCE SNe\label{tab:sample}}
\centering
\begin{tabular}{crrcccc}
\hline
Name & R.A. & Dec. & Redshift $^{\rm a}$ & $E(B-V)_{\rm MW}$$^{\rm b}$ & $r_{\rm peak}$ Magnitude$^{\rm c}$ & Host Distance \\
     &      &      &                     & (mag)                       & (mag)                              & (arcsec) \\
\hline
WFST-PS240307bn & 11h14m18.0s & -00d20m28.2s & 0.039 & 0.031 & -16.76$\,\pm\,$0.01 & 2.71\\
WFST-PS240307bh & 14h14m15.0s & +04d30m46.9s & 0.076 & 0.022 & -19.02$\,\pm\,$0.01 & 0.29\\
WFST-PS240503d  & 13h39m36.6s & +04d45m39.9s & 0.085 & 0.026 & -17.25$\,\pm\,$0.15 & 2.74\\
WFST250521xflp  & 13h09m03.2s & -00d03m07.6s & 0.081 & 0.021 & -16.66$\,\pm\,$0.03 & 5.40\\
WFST250522otkg  & 11h03m35.4s & +03d20m00.6s & 0.053 & 0.054 & -15.95$\,\pm\,$0.05 & 14.17\\
WFST250605fjov  & 13h30m11.5s & +01d51m41.9s & 0.083 & 0.022 & -17.28$\,\pm\,$0.04 & 2.08\\
WFST250617iqnc  & 13h17m05.3s & +01d49m46.4s & 0.036 & 0.031 & -15.47$\,\pm\,$0.04 & 3.78\\
\hline
\end{tabular}
\parbox{\linewidth}{\footnotesize
$^{\rm a}$ Spectroscopic redshift of the host galaxy described in \autoref{subsec:host}. \\
$^{\rm b}$ Galactic extinction derived from dust maps provided by \cite{schlafly_measuring_2011}. \\
$^{\rm c}$ $r$\text{-}band peak apparent magnitude of the main peak.
}
\end{table*}

\section{observations and data reduction} \label{sec:obs}
 The Wide Field Survey Telescope (WFST; \citealt{wang_science_2023})\footnote{\url{https://wfst.ustc.edu.cn}} is a 2.5 m optical telescope with a 6.5 deg$^2$ field of view and high \(u\)\text{-}band sensitivity, jointly developed by the University of Science and Technology of China (USTC) and the Purple Mountain Observatory (PMO). Its design enables deep, wide surveys of the northern sky, making it well suited for probing the dynamic universe and discovering extragalactic transients such as supernovae (SNe) and tidal disruption events (TDEs). WFST commenced a pilot survey (``WFST-PS") from Mar 4th to July 10, 2024, and started the six-year time-domain survey on 2024 December 14.

A total of 2800 SN candidates have been discovered by the Deep High-cadence $ugr$-band Survey project (``DH$ugr$;" J. Jiang et al. 2025, in prep), a key project for both pilot and formal surveys, as of 2025 June 30. The primary observing strategy involved daily/hourly-cadence photometries in $u$, $g$, and $r$ bands\footnote{However, due to an unexpected technical issue of the filter-exchange system occurred in late Mar 2024, only $g$- and $r$-bands data were obtained during the remaining WFST-PS period.}. The photometric data was derived using WFST data pipeline, based on the Large Synoptic Survey Telescope (LSST) soft ware stack \citep{axelrod_open_2010, bosch_hyper_2018, ivezic_lsst_2019}. Host galaxy spectra are available for approximately 400 candidates, the majority of the sample relies on photometric redshift determinations. WFST provided initial detections for most targets.

All 2800 light curves underwent comprehensive visual inspection and 56 candidates were identified with clear double-peaked signature. These early peaks likely originate from shock-cooling emission, such features suggest extended progenitor envelopes or interactions with circumstellar material. The following selection criteria were subsequently applied to this subset:
\begin{enumerate}[label=(\arabic*)] 
\item A minimum of five photometric measurements is required. These data must be obtained within 10 days of the initial detection. 
\item A distinct fading phase must precede the secondary rebrightening. Measurements during the decline were verified manually. Science images were inspected to ensure data validity. 
\item A reliable host spectroscopic redshift is mandatory for all candidates. 
\item Early light curve evolution must be detected in at least two bands. This includes the initial rise or decline. 
\end{enumerate}

Seven SNe are ultimately selected as the final sample for detailed analysis. The properties of all samples are shown in \autoref{tab:sample}.

Direct spectroscopic classification is precluded by the low apparent magnitudes of the samples. This limitation persists even at peak phases (fainter than 20 mag in most SNe). Consequently, all redshifts are derived from host galaxy spectra provided by DESI \citep{collaboration_data_2025}. Non-detection constraints are absent for WFST-PS240307bn and WFST250521xflp. These targets were detected during the commencement of WFST operations. Peak magnitudes in the $r$\text{-}band were derived via Gaussian Process regression and the full light curve was utilized for this fitting. Discovery magnitudes of all SNe are in the $g$\text{-}band range primarily from 19.3 to 22.0 mag.

\subsection{Extinction Correction}
Interstellar dust extinction affects the observed properties of SNe; therefore, accurate extinction correction is essential for linking observations with theoretical models. The total extinction is composed of two components: Galactic extinction from the Milky Way and host-galaxy extinction arising from the local environment of the SN. Galactic extinction is corrected using the reddening maps of \cite{schlafly_measuring_2011}, with extinction coefficients calculated according to the extinction law of \cite{cardelli_relationship_1989}. We adopt a standard interstellar medium extinction model with $R_V = 3.1$ throughout this work.

Spectroscopic observations are unavailable for the current sample. Consequently, host-galaxy reddening is estimated by comparing SN colors with templates, see \autoref{subsec:comparison}. Comparison indicates that host-galaxy reddening is minimal. Furthermore, most SNe are located at large offsets ($~>~2^{''}$) from their host centers. Host galaxy extinction is therefore neglected in this analysis.

\subsection{Light Curve}

\begin{figure*}
    \centering
    \includegraphics[width=0.7\textwidth]{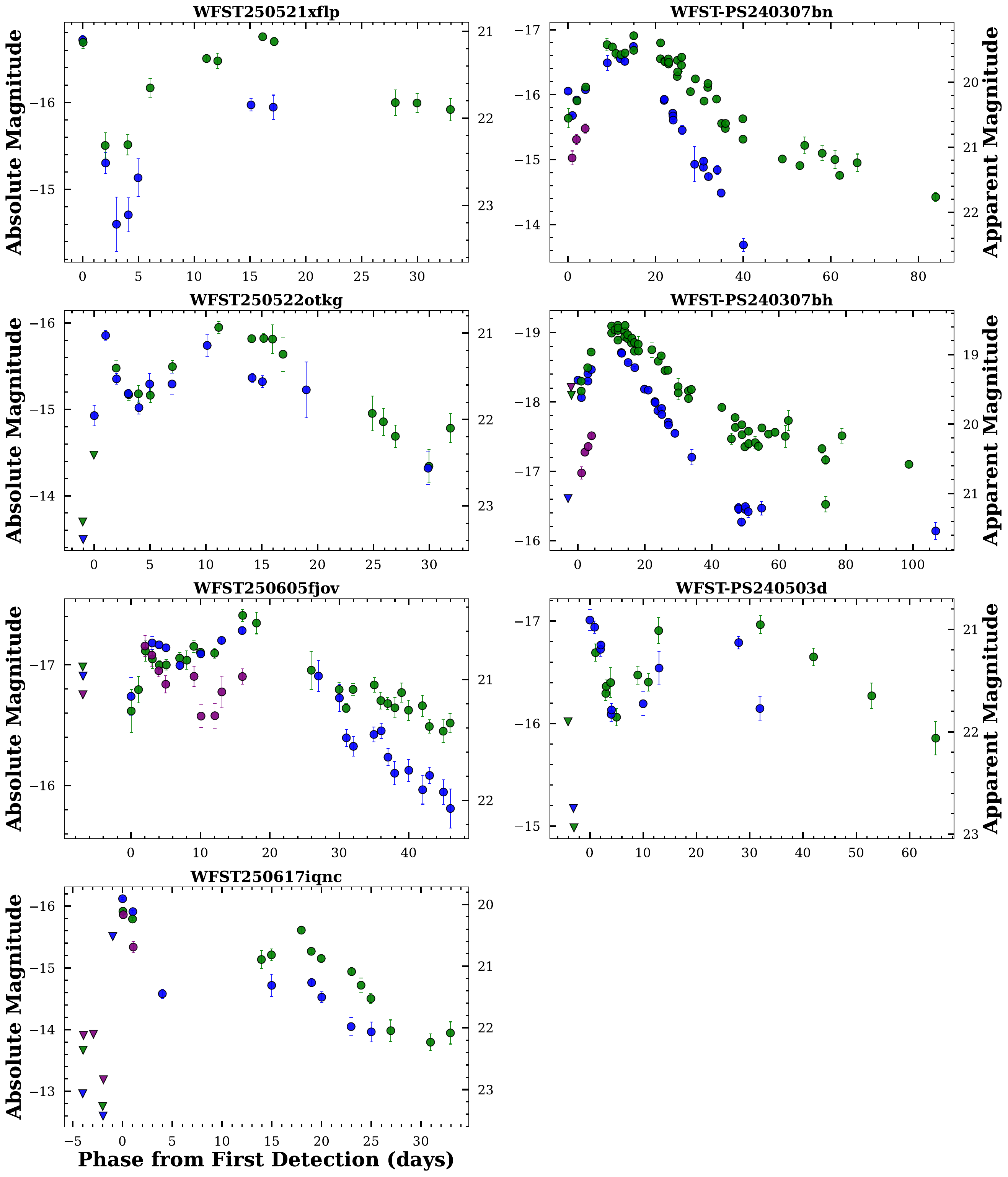}
    \caption{Multiband light curves are presented for the full sample. The $g$-band data are shown in blue, green and purple symbols denote $r$\text{-} and $u$\text{-}bands, respectively. Triangles indicate $3\sigma$ upper limits. Non-detections separated by $>7$ days from the first detection are omitted.}
    \label{fig:light curve}
\end{figure*}

Multiband light curves for the entire sample are presented in \autoref{fig:light curve}, where data are corrected for extinction. Available $3\sigma$ non-detections are included. 

The observational frame is used in the figure, and a time dilation effect correction is applied to the absolute magnitude as follows:
\begin{equation}
M_{\text{corr}} = M + 2.5 \times \log_{10}(1+z).
\end{equation}
Here $z$ denotes the redshift, accounting for corrections due to cosmological time dilation. A flat $\Lambda$CDM cosmology is adopted throughout this work. The Hubble constant is set to $H_0 = 67.66$ km s$^{-1}$ Mpc$^{-1}$, and the matter density parameter is $\Omega_m = 0.3111$. These values are utilized for all luminosity distance calculations and corrections.

The sample SNe show a rapid early decline within days of discovery followed by a luminous second peak (decline rates in $g$\text{-} and $r$\text{-}bands were derived for the seven sources over a $\sim3$ day rest-frame period; \autoref{tab:decline}). Notably, WFST250522otkg and WFST250605fjov were detected prior to the first maximum. These observations trace the surface heating and expansion driven by the shock wave propagating through the hydrogen envelope, as well as the subsequent spectral evolution where the cooling blackbody peak shifts through the observed bands. WFST-PS240307bh lacks dense sampling during the cooling phase, but its tight non-detection limits effectively constrain the early evolution.

\begin{table*}
\caption{Shock cooling decline rate in $r$\text{-} and $g$\text{-}band of WFST SCE SNe\label{tab:decline}}
\centering
\begin{tabular}{lcccc} 
\hline
Name & Decline Start & Decline End & $\Delta m$ & Decline Rate \\
     &      (MJD)    & (MJD)       &  (mag)     &(mag d$^{-1}$)\\
\hline
WFST-PS240503d   & 60432.74 & 60436.80 & $0.88\,\pm\,0.12$ & $0.23\,\pm\,0.03$ \\
WFST250521xflp   & 60725.76 & 60728.79 & $2.12\,\pm\,0.32$ & $0.76\,\pm\,0.11$ \\
WFST250522otkg   & 60727.72 & 60730.70 & $0.84\,\pm\,0.09$ & $0.30\,\pm\,0.03$ \\
WFST250605fjov   & 60812.71 & 60816.71 & $0.19\,\pm\,0.06$ & $0.05\,\pm\,0.02$ \\
WFST250617iqnc   & 60821.67 & 60825.67 & $1.54\,\pm\,0.07$ & $0.40\,\pm\,0.02$ \\
WFST-PS240307bh  & 60386.83 & 60387.86 & $0.25\,\pm\,0.03$ & $0.26\,\pm\,0.03$ \\
WFST-PS240307bn  & 60375.70 & 60376.70 & $0.38\,\pm\,0.04$ & $0.39\,\pm\,0.04$ \\
\hline
\end{tabular}
\\[10pt]
\parbox{\linewidth}{\footnotesize
\textbf{Note:} WFST-PS240307bh and WFST-PS240307bn possess fewer than two $r$-band detections during shock cooling. Consequently, $g$\text{-}band light curves are utilized for decline rate determinations. Values are derived directly from the photometric measurements.
}
\end{table*}

WFST250521xflp shows the steepest decline with a magnitude drop of $2.12\,\pm\,0.32$ mag over 2.80 days, corresponding to a rate of $0.76\,\pm\,0.13$ mag d$^{-1}$. In contrast, WFST250605fjov shows the slowest evolution with a rate of approximately $0.05\,\pm\,0.02$ mag day$^{-1}$. 

For comparison, SN 2016gkg features a $V(R)$\text{-}band decline rate of $0.68\,\pm\,0.06$ ($0.66\,\pm\,0.10$) mag d$^{-1}$ \citep{bersten_surge_2018}. This value falls within the range of the current sample, suggesting a common physical origin for the emission. Late-time evolution is generally unobserved due to the faintness and high redshift of the sources. Consequently, the $^{56}$Ni-powered radioactive tail remains unconstrained. However, this absence does not affect the analysis of the primary shock-cooling mechanism.

\subsection{Host} \label{subsec:host}
Host spectra are obtained from Data Release 1 of the Dark Energy Spectroscopic Instrument (DESI-DR1; \citealt{collaboration_data_2025}). DESI is a large-scale redshift survey targeting galaxies and quasars over ~14000 deg$^2$ of the extragalactic sky, provides high-precision redshifts that enable robust host galaxy identification and distance measurements for SNe. Besides host galaxies are matched from Data Release 16 of the Sloan Digital Sky Survey (SDSS-DR16), the fourth data release of the fourth phase of the Sloan Digital Sky Survey (SDSS-IV; \citealt{blanton_sloan_2017}). A positional cross-match are performed using coordinates of SNe to identify the nearest galaxy in projection and obtained their spectrums for further analysis. A total of 19 host galaxy spectra are secured, with at least one per SN. Multiple spectra for the same source show consistent redshifts and similar spectral morphologies. 

Host spectra are fitted using the Python package Bayesian Analysis of Galaxies for Physical Inference and Parameter EStimation (BAGPIPES\footnote{\url{https://bagpipes.readthedocs.io}}; \citealt{carnall_inferring_2018})  This package performs Bayesian spectral energy distribution (SED) analysis. The \texttt{MultiNest} algorithm is employed for parameter space exploration \citep{feroz_multinest_2009}. It provides posterior distributions and Bayesian evidence. Further modeling details are in \cite{carnall_vandels_2019}. Best-fit results are presented in \autoref{tab:host}.
\begin{table*}
\caption{Host Properties of WFST SCE SNe\label{tab:host}}
\centering
\begin{tabular}{ccccccc}
\hline
Name & log(SFR) & log(Stellar Mass) & $A_v$ & Age & Metallicity \\
     & ($M_{\odot}\,{\rm yr}^{-1}$) & ($M_{\odot}$) & (mag) & (Gyr) & ($Z_{\odot}$) \\
\hline
WFST-PS240307bn & $-2.27^{+0.03}_{-0.03}$ & $8.05^{+0.04}_{-0.04}$ & $0.11^{+0.07}_{-0.07}$ & $2.63^{+0.66}_{-0.50}$ & $0.12^{+0.02}_{-0.02}$ \\
WFST-PS240307bh & $-0.88^{+0.01}_{-0.01}$ & $9.77^{+0.00}_{-0.00}$ & $0.54^{+0.00}_{-0.00}$ & $1.72^{+0.01}_{-0.00}$ & $1.55^{+0.02}_{-0.02}$ \\
WFST-PS240503d  & $-0.68^{+0.01}_{-0.01}$ & $8.96^{+0.01}_{-0.01}$ & $0.05^{+0.06}_{-0.03}$ & $1.24^{+0.06}_{-0.06}$ & $1.66^{+0.08}_{-0.06}$ \\
WFST250522otkg  & $-0.40^{+0.00}_{-0.01}$ & $10.27^{+0.00}_{-0.00}$ & $0.83^{+0.01}_{-0.01}$ & $2.70^{+0.06}_{-0.06}$ & $1.90^{+0.03}_{-0.03}$ \\
WFST250521xflp  & $-0.58^{+0.00}_{-0.00}$ & $9.47^{+0.00}_{-0.00}$ & $0.51^{+0.02}_{-0.02}$ & $1.32^{+0.04}_{-0.02}$ & $1.90^{+0.05}_{-0.05}$ \\
WFST250605fjov  & $-1.54^{+0.02}_{-0.03}$ & $8.43^{+0.04}_{-0.02}$ & $0.10^{+0.07}_{-0.06}$ & $1.54^{+0.27}_{-0.20}$ & $0.16^{+0.01}_{-0.01}$ \\
WFST250617iqnc  & $-0.73^{+0.00}_{-0.01}$ & $9.42^{+0.01}_{-0.01}$ & $0.76^{+0.03}_{-0.02}$ & $3.30^{+0.22}_{-0.20}$ & $2.29^{+0.02}_{-0.02}$ \\
\hline
\end{tabular}
\end{table*}

\section{analysis} \label{sec:analysis}

\subsection{Photometric Comparisons}\label{subsec:comparison}

\begin{figure}
    \centering
    \includegraphics[width=0.45\textwidth]{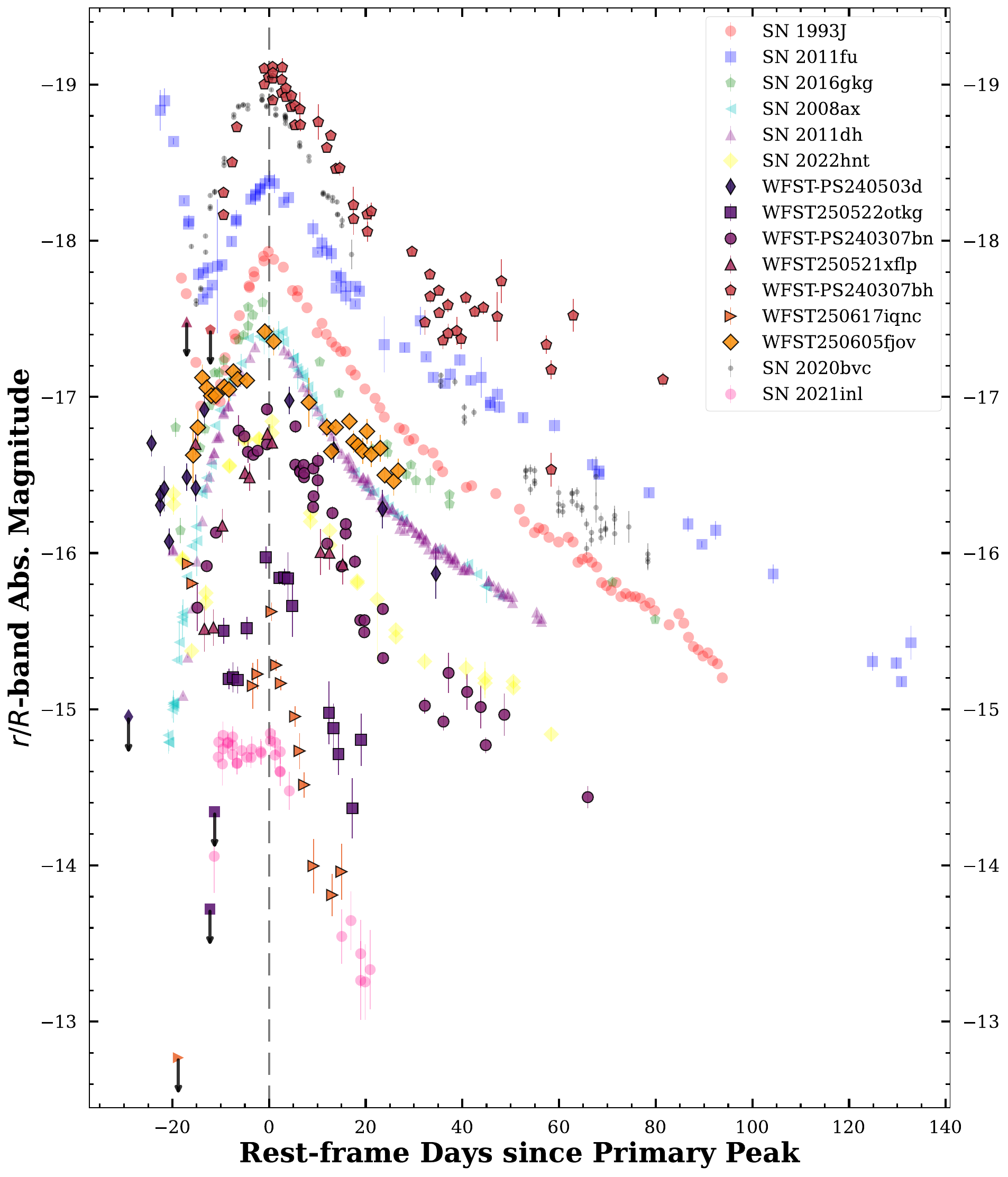}
    \caption{Light curves in $r/R$\text{-}band absolute magnitude of all SNe. For comparison, light curves of several well-studied SNe are plotted. Gray dashed line marks the phase of primary peak.}
    \label{fig:light curve comparation}
\end{figure}

\autoref{fig:light curve comparation} shows the comparison of absolute $r$- and $R$- band light curves between sample SNe and other well studied SESNe. 

WFST-PS240307bh represents the brightest object across all evolutionary phases. It shows an $r$\text{-}band peak absolute magnitude of $-19.02$, consistent with the Type Ib/c SN 2020bvc. Their light curve morphologies align closely from $-10$ to $40$ days. However, SN 2020bvc presents a faster decline after 50 days, likely attributable to UV leakage. In addition to WFST-PS240307bh, the brightness of the first peak is comparable or even higher than that of the second peak for the other six SNe. We note that although the second peak of WFST-PS240307bh is comparable to the peak of normal Type Ia supernovae (SNe Ia), the brightness of $< -18.2$ mag for the first $g$-band peak is too high for any early-excess scenario of SNe Ia \citep{jiang_hybrid_2017, jiang_surface_2018, maeda_type_2018}.
WFST250605fjov and WFST-PS240503d resemble SN 2016gkg and SN 2011dh around the primary peak. Their peak magnitudes are $-17.32$ and $-16.99$, respectively. These values align with SN 2016gkg ($-17.56$) and SN 2011dh ($-17.34$) in $R$ band.

WFST-PS240503d resembles SN 2016gkg during the shock-cooling phase. Conversely, WFST250605fjov displays a shorter interval between the first and primary peaks, indicating a more rapid evolution. WFST250521xflp displays brightness and evolution comparable to SN 2022hnt when excluding the shock-cooling phase. Notably, WFST250521xflp shows the steepest decline during shock cooling, likely stems from low envelope and ejecta masses.

\begin{figure}
    \centering
    \includegraphics[width=0.45\textwidth]{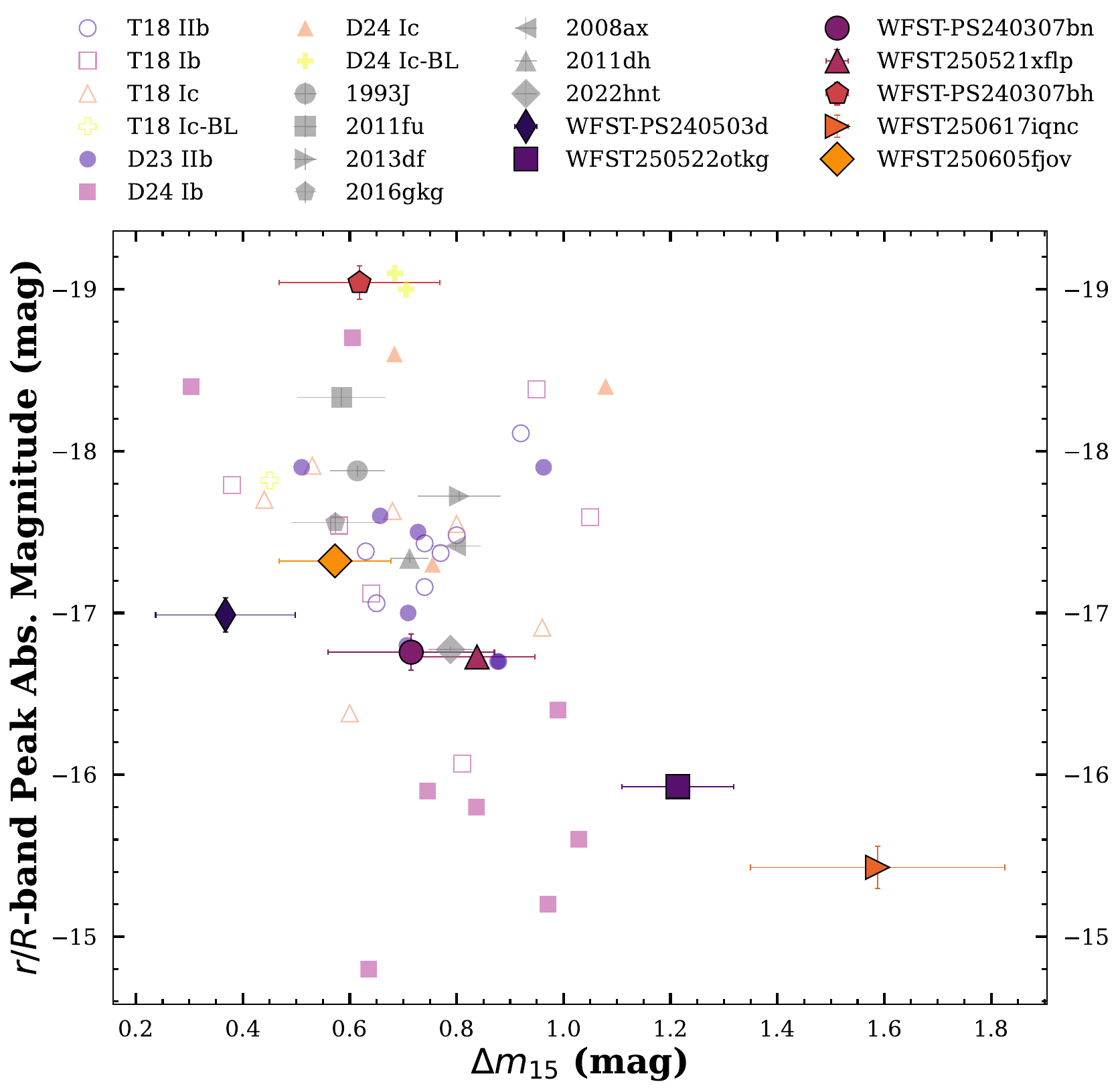}
    \caption{The light-curve decline-rate $\Delta m_{15}$ is plotted against peak brightness. Values correspond to the $r$\text{-} or $R$\text{-}band. The $R$\text{-}band is adopted for SN 1993J, SN 2011fu and SN 2013df. SN 2016gkg, SN 2008ax and SN 2011dh also utilize this band. All other sources rely on $r$\text{-}band observations. Comparison samples include SESNe from \cite{taddia_carnegie_2018}, D23 and D24. WFST-PS240307bh shows a peak brightness and decline rate comparable to Type Ic-BL SNe. Faster declining light curves generally possess lower peak luminosities. This trend associates high $\Delta m_{15}$ values with fainter magnitudes. WFST250522otkg and WFST250617iqnc display the highest decline rates. However, these measurements carry large uncertainties.}
    \label{fig:m15}
\end{figure}

WFST250522otkg and WFST250617iqnc represent the faintest SNe in the sample. Although SN 2021inl appears even fainter across all phases, it shows a slightly slower decline rate compared to the WFST targets, as shown in \autoref{fig:m15}. This discrepancy is attributed to redshift effects, as the observed band corresponds to bluer rest-frame wavelengths characterized by faster theoretical evolution.

\begin{figure*}
    \centering
    \includegraphics[width=0.7\textwidth]{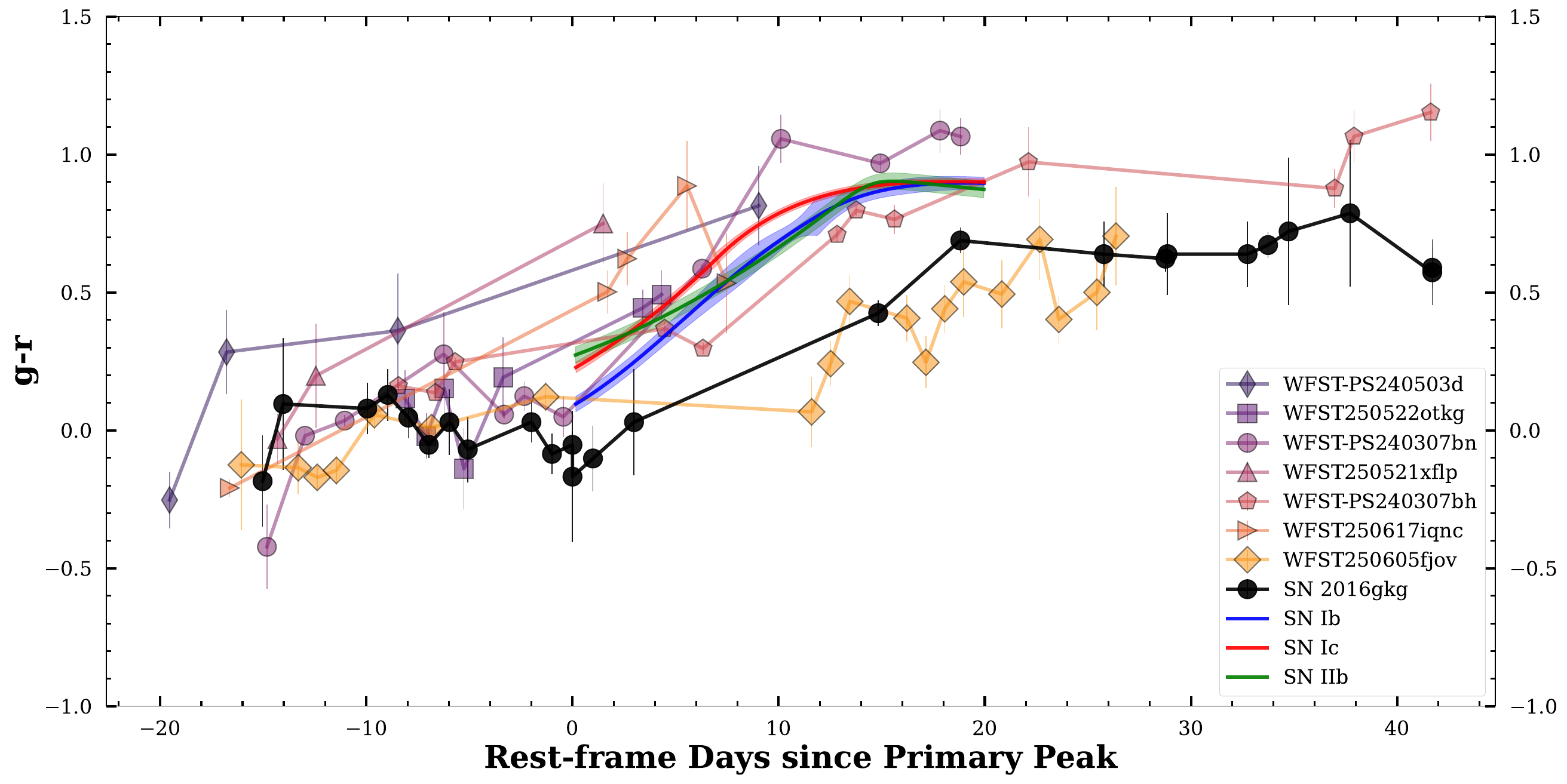}
    \caption{Color evolution of SNe compared to SN 2016gkg, where the $V-R$ color of SN 2016gkg is transformed to $g-r$ using the empirical color-color relation calibrated by \cite{jordi_empirical_2006}. Solid line with error range coded by colors are color evolution templates from the main peak to 20 days after in rest-frame of type IIb, Ib and Ic SNe introduced in \cite{stritzinger_carnegie_2018}.}
    \label{fig:color}
\end{figure*}

$g-r$ color evolution between WFST SN sample and SN 2016gkg is shown in \autoref{fig:color}. While some targets have limited multiband observations, resulting in fewer data points, the overall color trends are broadly consistent with SN 2016gkg. 

During the early phases, the $g-r$ color typically evolves from blue to red due to shock cooling. However, WFST0605fjov initially presents a blueward trend before rapidly reddening, indicating a temperature increase during its early phase. The $g-r$ color then undergoes another red-to-blue transform because of the second peak powered by the dacay of $^{56}$Ni subsequently turns to the red tail.

The $g-r$ color templates are also shown in figure, which represent Type IIb, Ib, and Ic SNe. They were derived from comprehensive statistical studies of SESN light curves, which show remarkably uniform color evolution during the early post-maximum phase \citep{stritzinger_carnegie_2018}. The uniformity spans 0 to +20 days relative to the $B$\text{-}band maximum. The overall colors of SNe are consistent with the templates.

\subsection{Bolometric Light Curve}
Bolometric light curves were derived using \texttt{SuperBol} \citep{nicholl_superbol_2018}, which accepts extinction-corrected apparent magnitudes. Pseudobolometric fluxes were computed by fitting blackbody spectral energy distributions (SEDs), integrating the flux over the full SED at each epoch.

Since detections are primarily limited to $g$\text{-} and $r$\text{-}bands, the lack of multiband constraints introduces potential uncertainties. High-order polynomial interpolations were applied to mitigate this issue, with the fitting order dependent on temporal coverage and data quality. $g$\text{-}band light curves were interpolated to the reference epochs defined by the better-sampled $r$\text{-}band.

\begin{figure*}
    \begin{minipage}[c]{0.45\textwidth}
        \centering
        \includegraphics[width=\textwidth]{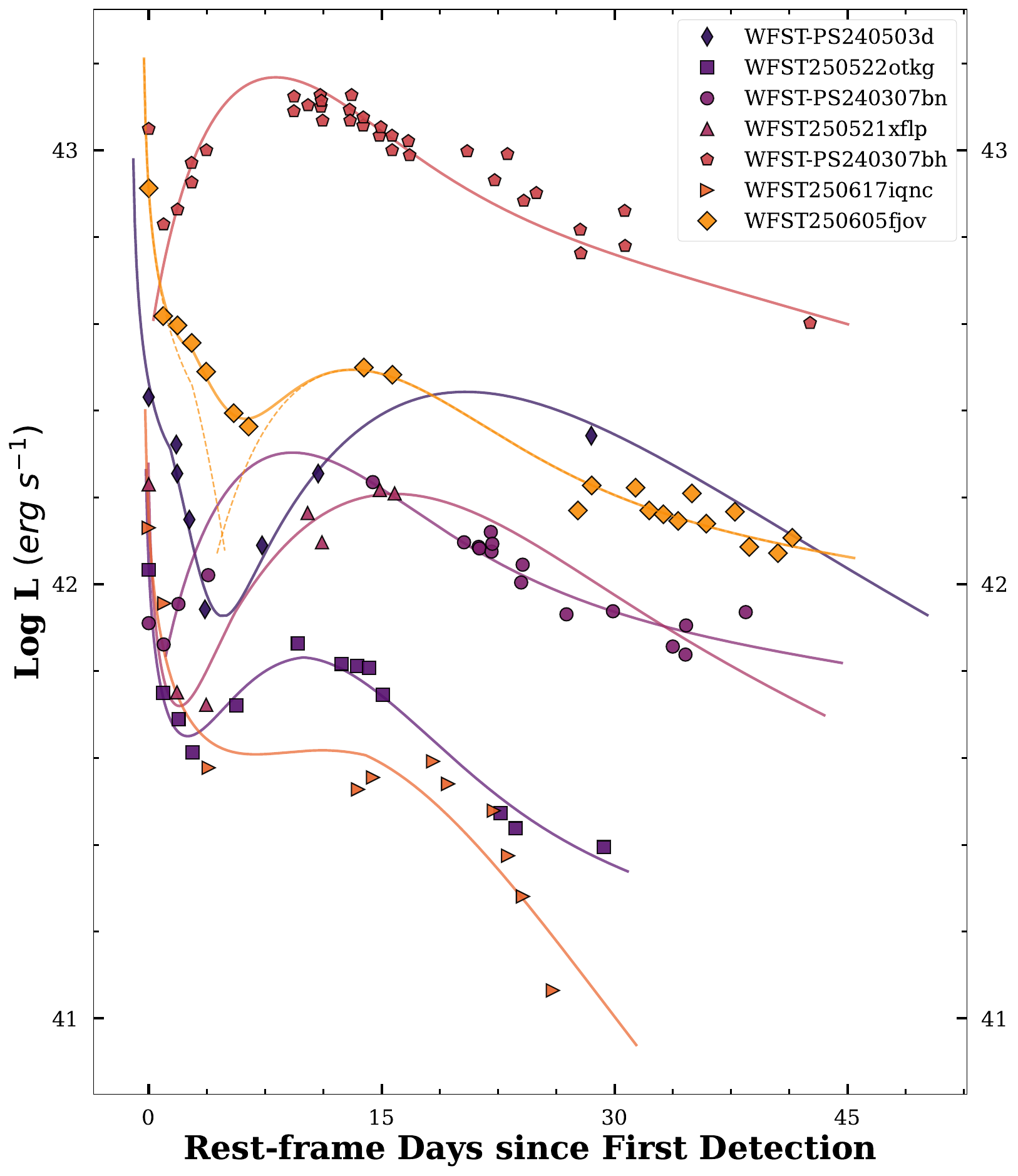}
    \end{minipage}
    \hfill 
    \centering
    \begin{minipage}[c]{0.45\textwidth}
        \centering
        \includegraphics[width=\textwidth]{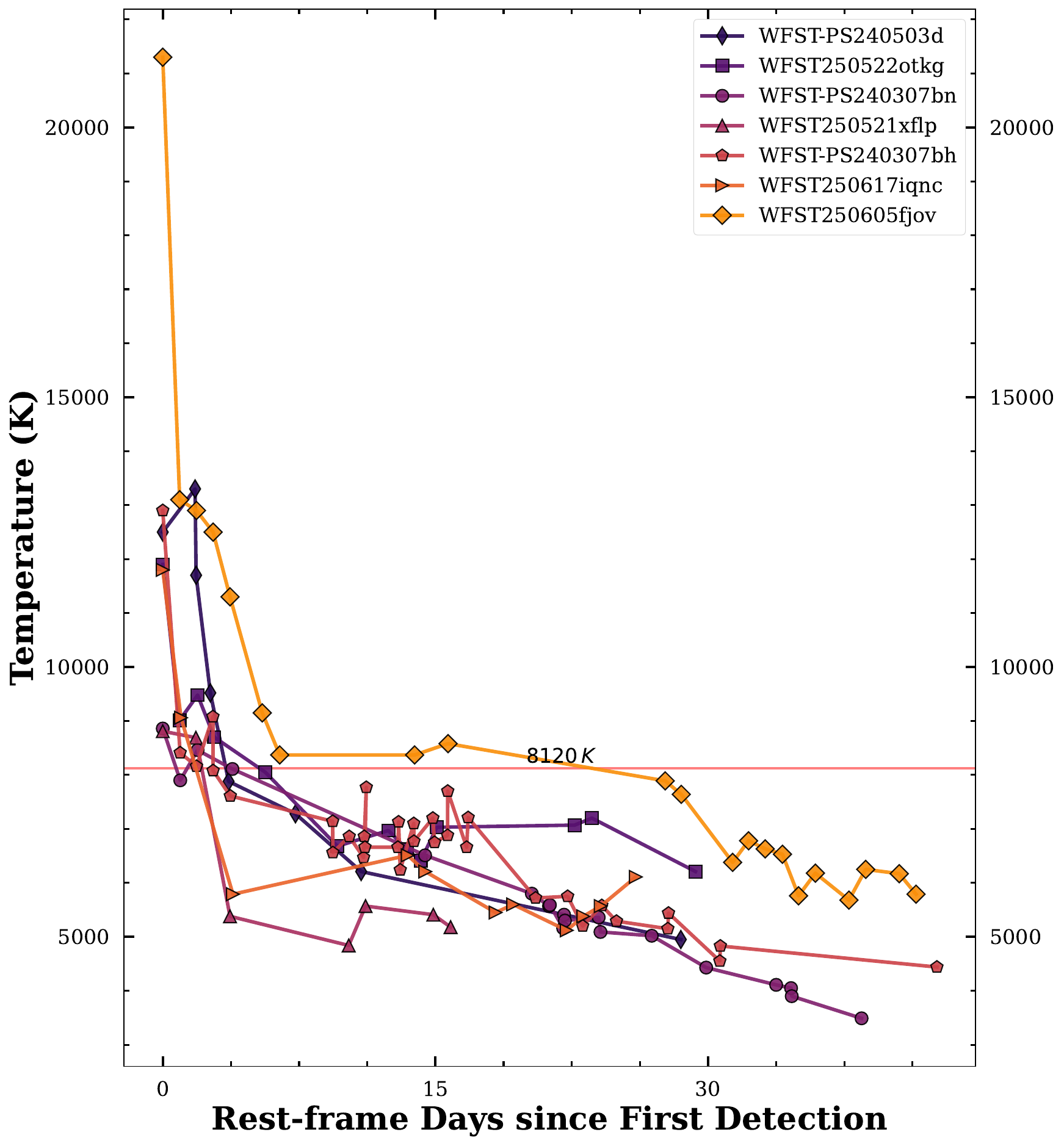}
    \end{minipage}
    
    \vspace{1em} 
    \caption{\textbf{Left}: The bolometric light curves of SNe. Solid lines denote the total best fits from Arnett and P21 model, with dotted lines showing the decomposed components for a representative example.
    \textbf{Right}: Black body temperature evolution of SNe derived from \texttt{SuperBol} fitting. Shock Cooling models are valid only for data where the $T_{\rm BB}$ is greater than $8120~\mathrm{K}$ or 0.7 eV \citep{sapir_uvoptical_2017}.}
    \label{fig:L}
\end{figure*}

WFST-PS240307bh, WFST250605fjov and WFST-PS240307bn possess early-phase $u$\text{-}band detections, provide improved constraints on the SED fitting. Consequently, the analysis treated sources with and without $u$\text{-}band data separately. A blackbody SED was fitted for each epoch. Bolometric luminosities were derived alongside blackbody temperatures. The resulting bolometric light curves are presented in \autoref{fig:L}, temperature evolution is also shown.

A temporal validity cut is applied based on a temperature threshold of $8120~K$ ($0.7~eV$), following the criterion of \cite{sapir_uvoptical_2017}. Temperature fitting errors are omitted from the figure because \texttt{SuperBol} cannot compute uncertainties with only two bands. Consequently, the independent approach of \cite{martinez_type_2022} is utilized to verify the results. This method employs empirical correlations between bolometric corrections (BCs) and colors, using $g-r$ to derive the necessary corrections.

The evolution is typically modeled in three phases: cooling, main peak, and tail. However, the current light curves do not extend to the tail phase, which usually begins $\sim 100$ days post-explosion. Therefore, this phase is excluded from the analysis. The remaining cooling and main peak phases are distinguished by the $8120~K$ boundary. Given the results from two methods are close, the interpolated data are utilized for the analysis.

\subsection{$^{56}$Ni-powered Light-Curve Modeling}
Physical parameters including the $^{56}$Ni mass, ejecta mass, and kinetic energy are derived from the radioactively powered second light-curve peaks. We employ the analytic model introduced by \citet{arnett_type_1982} to relate the observed bolometric light curves to the radioactive energy deposition from the \ce{^{56}Ni -> ^{56}Co -> ^{56}Fe} decay chain.

Although originally developed for Type Ia supernovae, this framework has been widely applied to stripped-envelope supernovae owing to the shared $^{56}$Ni-powered peak mechanism. Our approach is consistent with established applications in the literature \citep{taddia_carnegie_2018, lyman_bolometric_2016, dong_characterizing_2024}.

The model assumes homologous, spherically symmetric expansion with constant, time-independent optical opacity. The $^{56}$Ni distribution is treated as a centrally concentrated, point-like source with a negligible initial radius. Under these assumptions, the bolometric luminosity is expressed as a function of the ejecta mass $M_{\rm ej}$, kinetic energy $E_k$, and synthesized $^{56}$Ni mass $M_{\rm Ni}$.

\subsection{Shock-cooling Light-Curve Modeling}

Analytic models for shock-cooling emission differ primarily in their treatment of the envelope geometry and density structure. Representative formulations include those of \citet{sapir_uvoptical_2017} (hereafter SW17) and \citet{piro_shock_2021} (hereafter P21). The P21 model explicitly distinguishes between outer and inner envelope regions, highlighting a sensitivity to the polytropic index. In contrast, SW17 demonstrated that shock-cooling emission is relatively insensitive to this index and shows only a weak dependence on the progenitor's density structure. Initial tests using the SW17 prescription result in systematically poorer fits to our dataset. We therefore adopt the P21 model to describe the shock-cooling phase, while the secondary radioactive-powered peak is modeled using the Arnett formalism.

\begin{table*}
\caption{Best Parameters of Explosion from Fitting\label{tab:explosion parameters}}
\centering
\begin{tabular}{lccccc}
\hline
Name & $M_{\rm env}$ & $R_{\rm env}$ & $M_{\rm ej}$ & $M_{\rm Ni}$ & $E_k$ \\
     & ($M_{\odot}$) & ($R_{\odot}$) & ($M_{\odot}$) & ($M_{\odot}$) & ($10^{51}$ erg) \\
\hline
WFST-PS240503d   & $0.08^{+0.31}_{-0.02}$   & $122.47^{+230.13}_{-41.90}$ & $2.57^{+0.76}_{-0.41}$ & $0.22^{+0.03}_{-0.04}$ & $3.95^{+0.75}_{-1.27}$ \\
WFST-PS240307bn  & -                        & -                           & $1.17^{+0.63}_{-0.46}$ & $0.07^{+0.00}_{-0.00}$ & $0.86^{+1.53}_{-0.65}$ \\
WFST-PS240307bh  & -                        & -                           & $1.19^{+0.40}_{-0.49}$ & $0.49^{+0.00}_{-0.00}$ & $1.57^{+1.60}_{-1.11}$ \\
WFST250521xflp   & $0.33^{+0.09}_{-0.15}$   & $67.04^{+73.47}_{-25.17}$   & $1.54^{+0.69}_{-0.61}$ & $0.07^{+0.01}_{-0.01}$ & $2.80^{+1.69}_{-1.68}$ \\
WFST250522otkg   & $0.42^{+0.06}_{-0.13}$   & $257.24^{+153.32}_{-119.06}$& $1.09^{+0.46}_{-0.49}$ & $0.02^{+0.00}_{-0.00}$ & $2.18^{+1.73}_{-1.73}$ \\
WFST250605fjov   & $0.23^{+0.08}_{-0.06}$   & $188.61^{+180.87}_{-63.35}$ & $1.68^{+0.65}_{-0.58}$ & $0.12^{+0.00}_{-0.00}$ & $1.76^{+2.27}_{-1.11}$ \\
WFST250617iqnc   & $0.42^{+0.05}_{-0.09}$   & $300.25^{+134.90}_{-76.01}$ & $1.18^{+0.49}_{-0.21}$ & $0.01^{+0.00}_{-0.00}$ & $2.97^{+1.48}_{-1.37}$ \\
\hline
\end{tabular}
\parbox{0.8\linewidth}{\footnotesize
\textbf{Note.} Reliable constraints on $M_{\rm env}$ and $R_{\rm env}$ could not be derived for WFST-PS240307bn and WFST-PS240307bh. The extremely brief duration of the shock-cooling phase in these events precluded robust fitting.
}
\end{table*}

To describe the full double-peaked light curves, we employ a combined framework in which the total flux is computed as the sum of the shock-cooling and radioactive components. Although these components are often fitted independently in the literature \citep{subrayan_early_2025, chen_sn_2025}, we perform a simultaneous fit in order to reduce parameter degeneracies. A common optical opacity is enforced for both components, ensuring a physically self-consistent description of the light-curve evolution. The best-fit parameters are summarized in \autoref{tab:explosion parameters}, and the corresponding bolometric light-curve fits are shown in \autoref{fig:L}.

We note that more general modeling frameworks that explicitly solve the time-dependent radiative diffusion equation and self-consistently track the transition from shock-cooling emission to radioactive heating have recently been developed in the literature (e.g., \texttt{TransFit}; \citealt{Liu2025}). However, given the limited temporal coverage and signal-to-noise ratio (S/N) of the early-time data in our sample, we adopt the widely used analytic prescription of \citet{piro_shock_2021}, which captures the dominant physical dependencies while introducing only a minimal number of free parameters.

\begin{figure*}
    \centering
    \begin{minipage}[t]{0.48\textwidth}
        \centering
        \includegraphics[width=\linewidth]{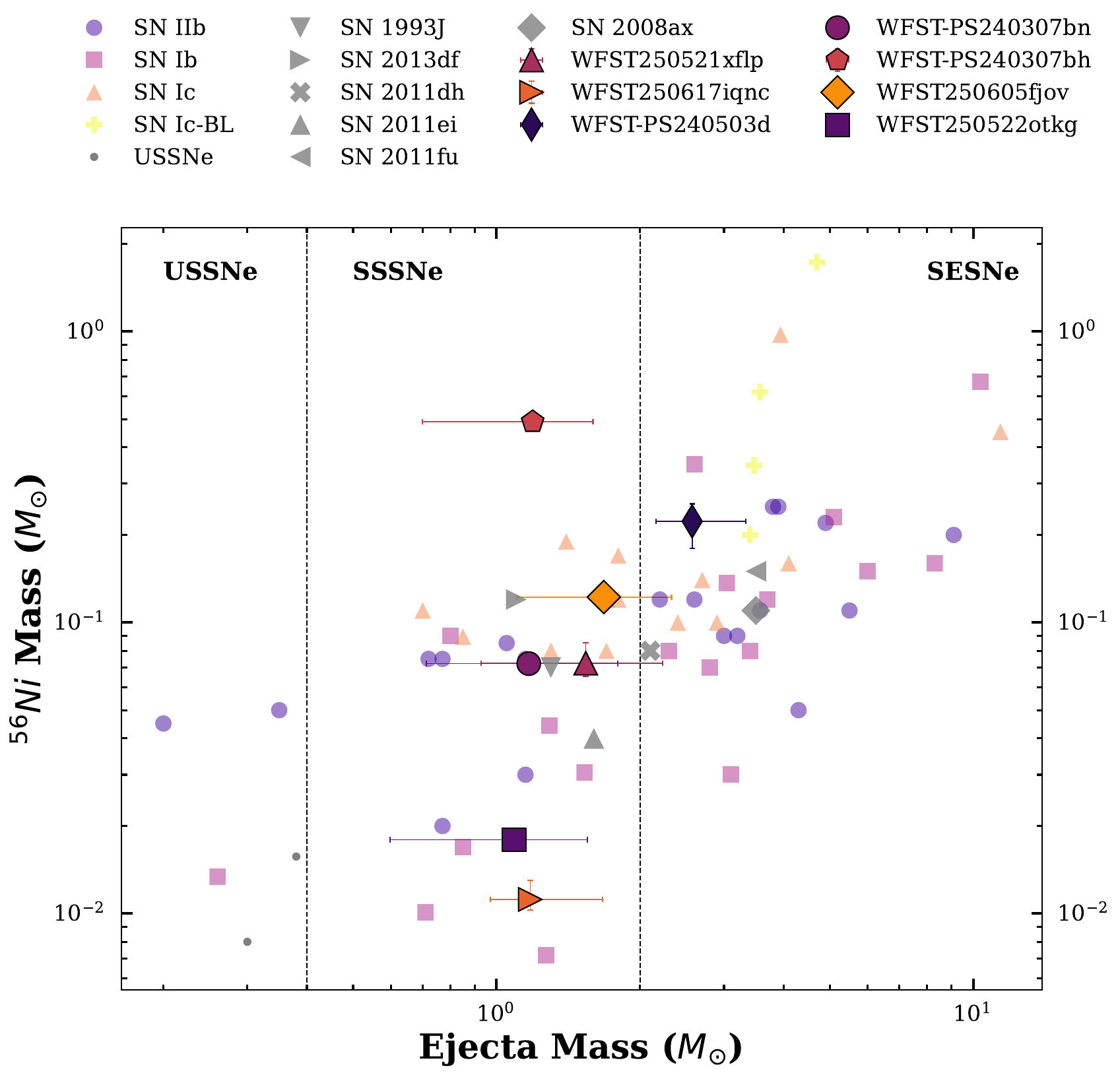}
        \captionof{figure}{Best fit of ejecta mass versus nickel mass of SNe from Arnett model, well studied type IIb SNe and SESNe are plotted for comparison.}
        \label{fig:Mej}
    \end{minipage}
    \hfill
    \begin{minipage}[t]{0.48\textwidth}
        \centering
        \includegraphics[width=\linewidth]{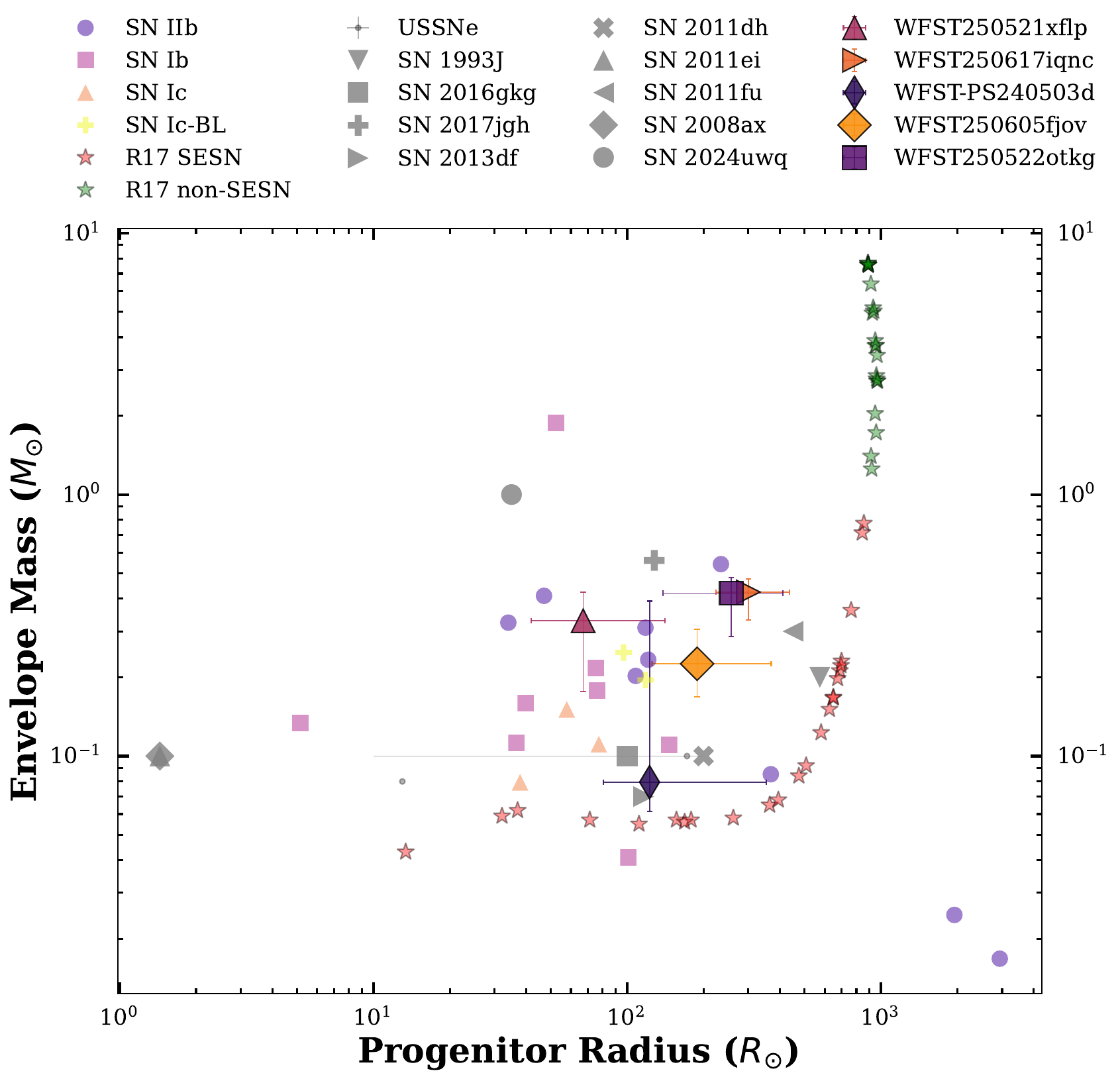}
        \captionof{figure}{Progenitor radius versus envelope mass of SNe, well studied type IIb SNe and SESNe with progenitor parameters are plotted for comparison.}
        \label{fig:menv}
    \end{minipage}
\end{figure*}

\subsection{Constraining Progenitor Luminosity}

\section{Discussion} \label{sec:discussion}
We place explosion and progenitor parameters derived from modeling in the context of well-studied SESNe to investigate correlations between key physical properties. This comparison allows us to assess the location of our sample in the parameter space and to probe for any distinctive characteristics that may shed light on progenitor diversity or explosion mechanisms.

\subsection{Physical Parameters}
Ejecta masses and $^{56}$Ni masses for the sample SNe are presented in \autoref{fig:Mej}, alongside comparative data for SESNe from \cite{taddia_carnegie_2018}, D23, and D24. Additionally, two ultra-stripped envelope SNe (USSNe) from \cite{de_iptf_2018} and \cite{yao_sn2019dge_2020} are included. 

Most events cluster within a moderate ejecta mass region of $0.4 - 2 M_{\odot}$, consistent with the transitional type between USSNe and SESNe described in D23, characterized by strongly stripped envelopes. WFST-PS240503d represents the sole exception, showing the highest ejecta mass of $\sim 2.6M_\odot$. 

Progenitor radii are plotted against envelope masses in \autoref{fig:menv}. The derived envelope masses range from 0.07 to 0.5 $M_\odot$, while radii span 60 to 300 $R_\odot$. These properties suggest Yellow Supergiant (YSG) or Blue Supergiant (BSG) progenitors, with hydrogen envelopes stripped via single-star winds or binary mass transfer. 

Theoretical predictions for SN IIb envelope masses and radii from \cite{ouchi_radii_2017} (hereafter R17) are also displayed for comparison; these were computed using the \texttt{MESA} code under a binary evolutionary channel assumption. At fixed envelope masses, the current sample consistently yields more compact progenitor radii compared to these predictions. This discrepancy is likely attributable to convection efficiency: as noted by \cite{dessart_type_2013}, efficient convection yields compact progenitors, whereas small mixing length parameters in models may result in overestimated radii. Additionally, the fact that 
R17 fixes the initial primary mass at 16 $M_{\odot}$ may account for the difference.

The mapping between helium core mass and zero-age main-sequence (ZAMS) mass depends sensitively on the choice of stellar evolution code, including \texttt{MESA} \citep{fang_inferring_2023, temaj_convective-core_2024}, \texttt{KEPLER} \citep{sukhbold_core-collapse_2016, sukhbold_high-resolution_2018}, and \texttt{HOSHI} \citep{takahashi_monotonicity_2023, fang_red_2025}. In contrast, the relationship between helium core mass and progenitor luminosity is found to be comparatively robust and largely independent of the adopted stellar evolution model \citep{fang_red_2025}. This consistency motivates the use of an approximately universal empirical relation of the form
\begin{equation}
\log\frac{L}{L_{\odot}} = 1.47 \times \log\frac{M_{\mathrm{He,core}}}{M_{\odot}} + 4.01.
\end{equation}

\begin{figure}
    \centering
    \includegraphics[width=0.45\textwidth]{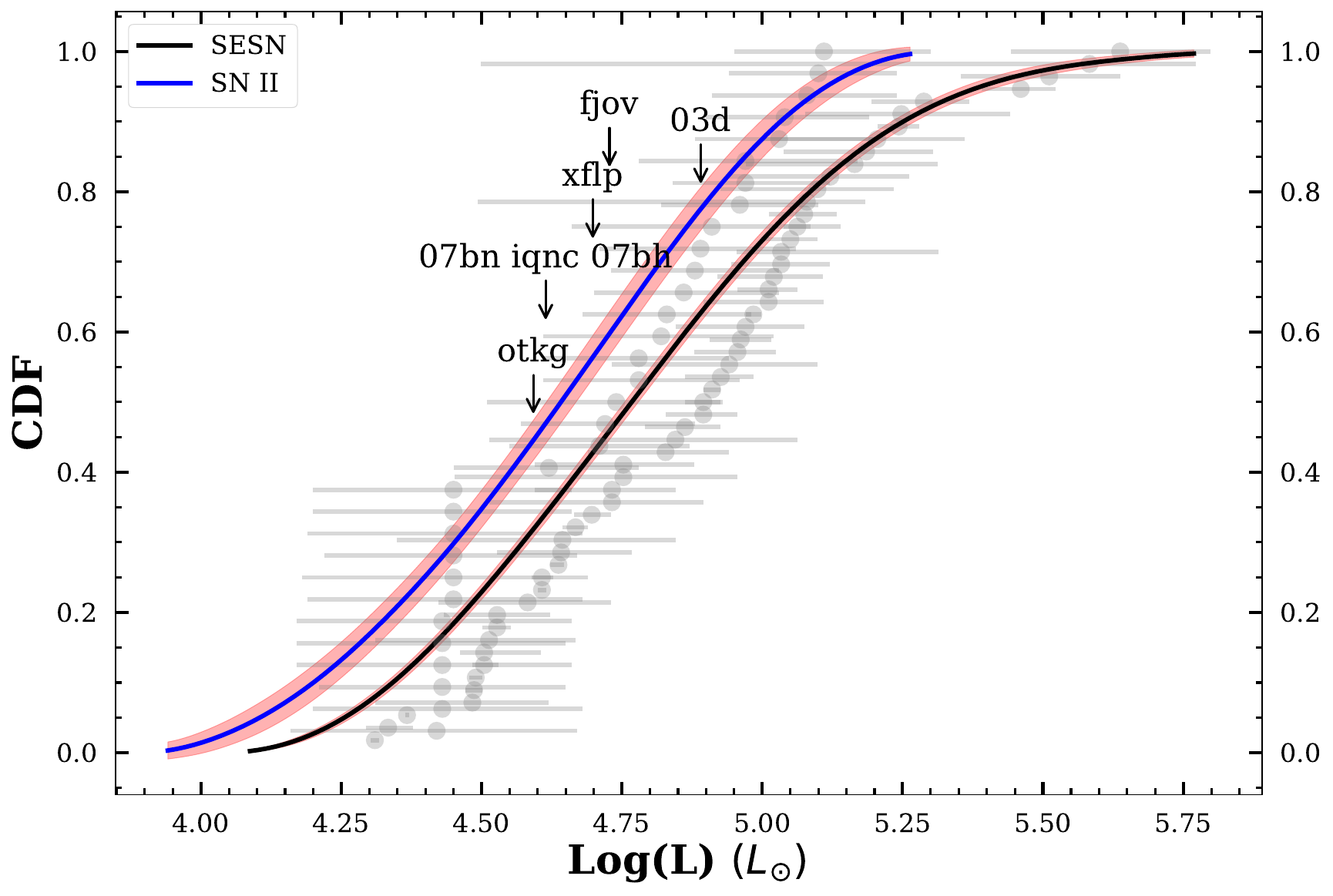}
    \caption{The cumulative distribution functions of the inferred progenitor luminosities for the SESNe samples from D23, D24, and \citet{taddia_carnegie_2018}. The grey circles in the background represent the individual data points, and the red shaded region indicates the $1\sigma$ error bounds. The luminosity positions of all samples are marked by black arrows.}
    \label{fig:Lpro}
\end{figure}

We assume a canonical compact remnant (neutron star) mass of $\sim 1.4\,M_{\odot}$ for all events. Under this assumption, helium core masses are estimated as $M_{\mathrm{He,core}} = M_{\rm ej} + 1.4\,M_{\odot}$. Progenitor luminosities are then inferred from the corresponding core masses using the above relation. The cumulative distribution functions (CDFs) of the inferred progenitor luminosities for the SESNe samples from D23, D24, and \citet{taddia_carnegie_2018} are shown in \autoref{fig:Lpro}.

The progenitor luminosity distribution for SNe II from \cite{fang_diversity_2025} is included for comparison; these values were constrained using nebular spectroscopy. To construct the CDF, luminosities were sampled from an asymmetric Gaussian distribution defined by literature means and uncertainties. This process was iterated 10,000 times to minimize statistical noise. Solid lines and shaded regions denote the median and $68\%$ confidence intervals, respectively. The positions of the current sample are marked with downward arrows, showing luminosities ranging from $\sim10^{4.6}$ to $\sim10^{4.9}$ $L_{\odot}$.

The distributions of SNe II and SESNe are similar at low luminosities ($L\lesssim10^{5.25}$ $L_{\odot}$), whereas SESNe extend to higher luminosities ($L\lesssim10^{5.75}$ $L_{\odot}$). The progenitor luminosities of the current sample fall between the SN II and SESN populations. This intermediate distribution is consistent with predictions for blue SN IIb progenitors ($\sim 10^{4.5}-10^{5.5}\,L_{\odot}$) that evolve primarily via binary channels \citep{sravan_progenitors_2020}.

\section{Conclusion} \label{sec:conclusion}

We present early multiwavelength photometric observations of seven supernovae discovered by WFST, all of which show prominent shock-cooling emission at early times. The main conclusions of this work are summarized as follows:

\begin{enumerate}[label=\arabic*.]

\item All events show a clear double-peaked light-curve morphology. The early excess emission is naturally interpreted as shock-cooling emission and typically persists for $\sim$3–5 days. This behavior closely resembles that observed in stripped-envelope supernovae (SESNe) such as SN~2016gkg and SN~1993J, indicating partially stripped progenitors. The brightest event, WFST-PS240307bn, reaches an $r$-band primary peak absolute magnitude of $-19.02$, likely consistent with a Type Ic classification, while the remaining events peak at $M_r \sim -15.5$ to $-17.3$, comparable to typical Type IIb supernovae.

\item By jointly modeling the shock-cooling and radioactive-powered phases using the P21 and Arnett frameworks, we derive physical constraints on the explosion and progenitor properties. The inferred ejecta masses ($\sim1.1$–$2.6\,M_{\odot}$) place these events in the transitional regime between ultra-stripped supernovae (USSNe) and normal SESNe. In addition, the modest envelope masses ($\sim0.1$–$0.4\,M_{\odot}$), together with progenitor radii of $\sim120$–$260\,R_{\odot}$ and luminosities of $10^{4.6}$–$10^{4.9}\,L_{\odot}$, imply extensive pre-explosion envelope stripping. Such substantial mass loss strongly favors a binary evolutionary origin.

\end{enumerate}

\begin{acknowledgments}

This work was supported by the Strategic Priority Research Program of the Chinese Academy of Sciences (Grant No. XDB0550300), the National Natural Science Foundation of China (NSFC; Grant Nos. 12393811 and 12303047), the National Key Research and Development Program of China (Grant Nos. 2023YFA1608100 and 2021YFA0718500), and the Natural Science Foundation of Hubei Province (Grant No. 2023AFB321). J.J. acknowledges support from the Japan Society for the Promotion of Science (JSPS) KAKENHI (Grant No. JP22K14069). K.M. acknowledges support from the Japan Society for the Promotion of Science (JSPS) KAKENHI grant (JP24KK0070, JP24H01810). The work is partly supported by the JSPS Open Partnership Bilateral Joint Research Projects between Japan and Finland (K.M and H.K; JPJSBP120229923). H.K. was funded by the Research Council of Finland projects 324504, 328898, and 353019. L.G. acknowledges financial support from CSIC, MCIN and AEI 10.13039/501100011033 under projects PID2023-151307NB-I00, PIE 20215AT016, and CEX2020-001058-M.

The Wide Field Survey Telescope (WFST) is a joint facility of the University of Science and Technology of China and Purple Mountain Observatory.

\end{acknowledgments}

\facilities{WFST}

\software{
    Astropy \citep{collaboration_astropy_2022}, pandas \citep{pandas_v3_2026}, numpy \citep{harris_array_2020}, scipy \citep{virtanen_scipy_2020}, Jupyter-notebook \citep{kluyver2016jupyter}, SWarp \citep{bertin2010swarp}, SExtractor \citep{1996A&AS..117..393B}, superbol \citep{nicholl_superbol_2018}, BAGPIPES \citep{carnall_inferring_2018}, astroquery \citep{ginsburg_astroquery_2019}
}

\bibliography{WFSTSCE}{}
\bibliographystyle{aasjournalv7}

\end{document}